\documentclass[10pt,twocolumn,letterpaper]{article}

\usepackage{cvpr}
\usepackage{times}
\usepackage{epsfig}
\usepackage{graphicx,float}
\usepackage{amsmath}
\usepackage{amssymb}
\usepackage{pgfplots}
\usepackage{pgfplotstable}
\usepackage{tikz}
\usepackage{array}
\usepackage{booktabs}
\usepackage{placeins}
\usepackage{color,soul}
\usepackage{caption}

\DeclareMathOperator*{\argmin}{arg\,min}
\newcommand{\boldhead}[1]{\vspace{0.02in}\noindent\textbf{#1.}}

\usepackage[pagebackref=true,breaklinks=true,letterpaper=true,colorlinks,bookmarks=false]{hyperref}

\cvprfinalcopy 

\def\cvprPaperID{****} 

\begin{document}

\title{Efficient Media Retrieval from Non-Cooperative Queries}

\author{Kevin Shih$^\dag$, Wei Di$^\ddag$, Vignesh Jagadeesh$^\ddag$, Robinson Piramuthu$^\ddag$\\
$^\dag$University of Illinois at Urbana-Champaign $^\ddag$eBay Research Labs\\
{\tt\small kjshih2@illinois.edu,[wedi, vjagadeesh, rpiramuthu]@ebay.com}}
\maketitle

\begin{abstract}
  Text is ubiquitous in the artificial world and easily attainable when it comes to book title and author names. Using the images from the book cover set from the Stanford Mobile Visual Search dataset and additional book covers and metadata from openlibrary.org, we construct a large scale book cover retrieval dataset, complete with 100K distractor covers and title and author strings for each. 

  Because our query images are poorly conditioned for clean text extraction, we propose a method for extracting a matching noisy and erroneous OCR readings and matching it against clean author and book title strings in a standard document look-up problem setup. Finally, we demonstrate how to use this text-matching as a feature in conjunction with popular retrieval features such as VLAD using a simple learning setup to achieve significant improvements in retrieval accuracy over that of either VLAD or the text alone.
\end{abstract}

\section{Introduction}
Large-scale image-based product look-up is an increasingly sought after feature as more people begin to make financial transactions through their mobile devices. In order for such a  feature to be practical, not only must it be accurate, but also able to return results within a matter of seconds for huge databases. The type of methods that best fit this bill are generally built around Bag-of-Words features as they are compatible with hash tables and approximate nearest neighbor approaches. In this work, we attempt to use text recognition techniques to treat image retrieval as an actual text document look-up problem.

In this work, we focus on the book cover based image retrieval. As is the case for many artificial products, there is a significant amount of informative textual information on almost every cover that would be extremely beneficial in the retrieval process. This idea has been applied successfully in book spine look-up~\cite{Tsai2011Spine}, but there is no existing study to our knowledge that tests this on a large-scale setting. Further, unlike~\cite{Tsai2011Spine}, we are interested in cases where the text is much harder to localize, such as mobile snapshots taken from suboptimal angles and lighting conditions. 

Recognizing the dearth of large-scale book cover retrieval datasets with text annotations, we create our own for this work. Using an existing retrieval dataset~\cite{chandrasekhar2011stanford} comprising mobile snapshots of book covers, we augment it with more than 100K additional distractor book cover images to emulate a large-scale use case. We then provide textual book cover and author information for all book covers, including for the distractors, which is not only readily available but also extremely helpful in the retrieval task as we will demonstrate.

In addition to providing a large-scale text-augmented dataset, we look at how to robustly use text in cluttered and poorly oriented images to match against clean text annotations.  While there is extensive work in detecting text in natural images, many have focused on performing well in popular datasets such as ICDAR '11~\cite{shahab2011icdar}, which feature mostly horizontal images. In our particular case of mobile book cover images, we expect text to appear at random orientations and viewing angles, as well as to have occasional occlusions. Running off-the-shelf OCR software directly on the image will generally yield no useful output as software such as Tesseract generally assumes text to be horizontal and well aligned. Using OCR software in conjunction with state-of-the-art text localization methods, we still expect the results to be riddled with transcription errors due to the awkward viewing angles, thereby ruling out any matching technique that relies on full words.  In our work, we demonstrate a method to robustly extract and use noisy and erroneous chunks of text from such images to match with clean text strings using methods based on approximate string matching.

Finally, we present a clean SVM-based formulation for combining multiple ranking signals such as text with popular retrieval features such as VLAD~\cite{jegou2010aggregating}. While the parameters for combining the signals can be easily determined with grid search, our use of rankSVM~\cite{Joachims2006RSVM} performs just as well and is straightforward to extend to any additional set of features.

In summary, our contributions are
\begin{itemize}
\item use of noisy OCR output as a visual feature in conjunction with traditional visual search techniques to get big performance gains for non-cooperative queries
\item seamless integration of multiple techniques using rankSVM
\item augmentation of dataset for large scale book retrieval with full text annotation.
\end{itemize}

\section{Related Work}

Popular datasets for image retrieval settings consist of a set of query images and their ground truth, however the set of query images are typically too small to emulate a large-scale setting. For large scale testing, millions of distractor images are usually added to the dataset. However, most of the current available datasets contain only natural scene images such as the INRIA  holiday dataset~\cite{jegou2008hamming}, Oxford buildings~\cite{philbin2007object}, and Zurich Building~\cite{shao2003zubud}. 

One notable product-retrieval dataset which we use to build our own is the Stanford Mobile Visual Search dataset (SMVS) proposed by  Chandrasekhar \etal \cite{chandrasekhar2011stanford}. This dataset contains smart-phone images of various products, CD covers, book covers and outdoor landmarks. For books, the reference images are clean catalogue images obtained from the product sites. The query images were taken indoors under varied lighting conditions with cluttered backgrounds.  While the query images depict a realistic scenario of product queries taken by the average smart-phone wielding consumer, the dataset itself still lacks in both scale as well as relevant textual annotations that we will be exploiting in this work. In~\ref{subsec:dataset}, we describe how we augment the book cover portion of this dataset to better emulate real-world product look-up  scenarios.
 
On the whole, literature on large-scale product retrieval, especially book covers, has been quite limited. One of the few relevant works is Matsushita \etal ~\cite{matsushita2011interactive} which introduces an interactive bookshelf system. The system includes cameras that can take pictures when a book is being stored or removed from the shelf, and uses the standard Speeded Up Robust Features (SURF)~\cite{bay2006Surf} feature to match the book to database, which obtain invariance to scale, illumination change, occlusion and rotation. However, the performance of such systems is not satisfactory as the exclusive use of such local image features as SURF, which are developed primarily for natural/wholistic images, does not leverage the rich text information~\cite{Tsai2011Spine} available in nearly every image instance. Other available studies on book products focus on recognizing books on shelves~\cite{Tsai2011Spine, chen2010building}. However, these approaches focus on book-spines which have easily localizable vertical text. Further, we are not aware of any large-scale extensions of these works.

\section{Method}
\begin{figure}[t]
  \scriptsize
  \begin{center}
    $
    \begin{array}{ccccc}
      \includegraphics[width=0.09\linewidth]{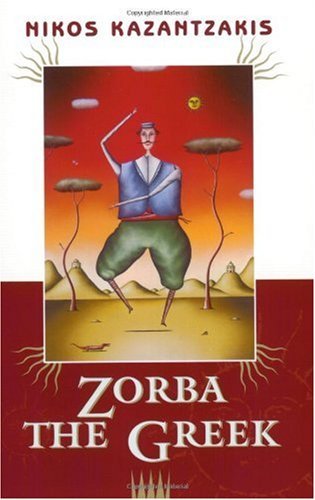} & \includegraphics[width=0.18\linewidth]{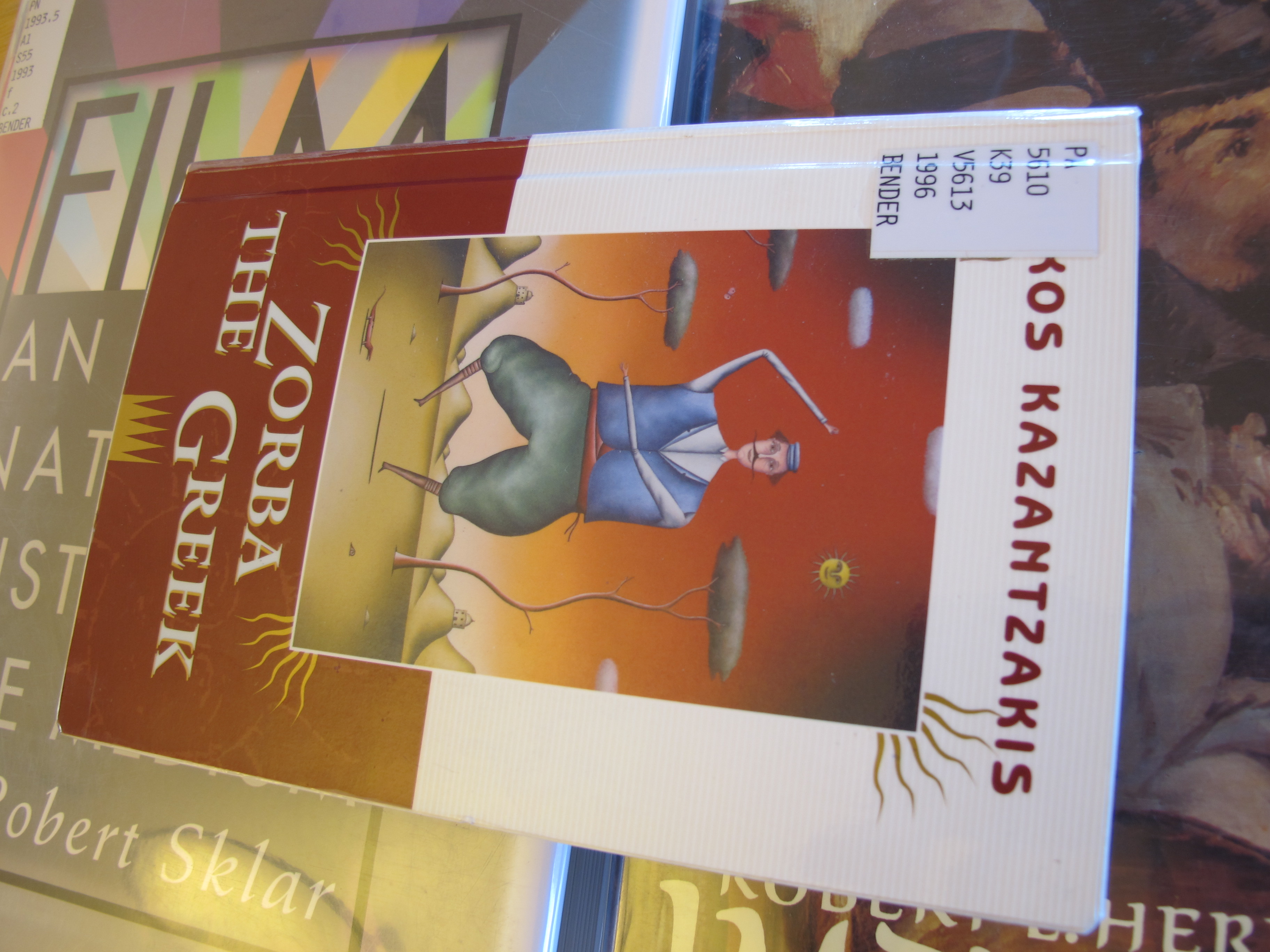} & \includegraphics[width=0.18\linewidth]{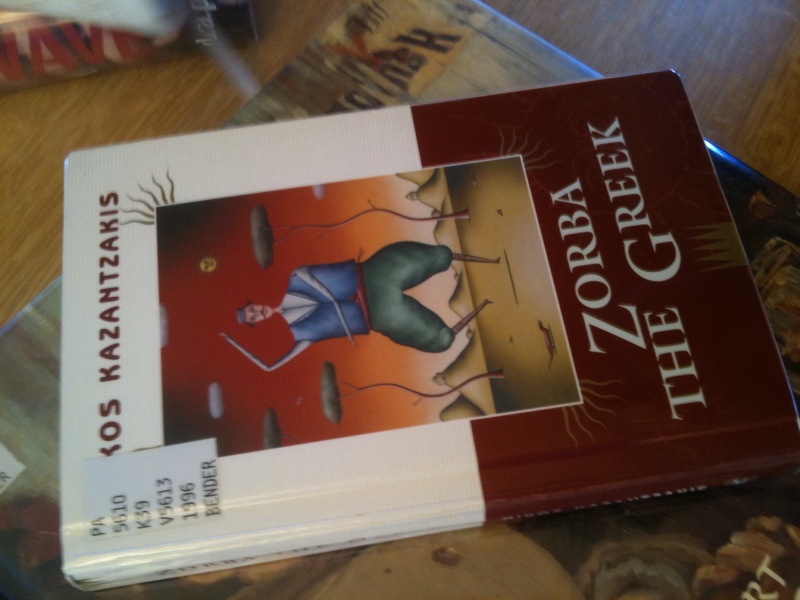} & \includegraphics[width=0.18\linewidth]{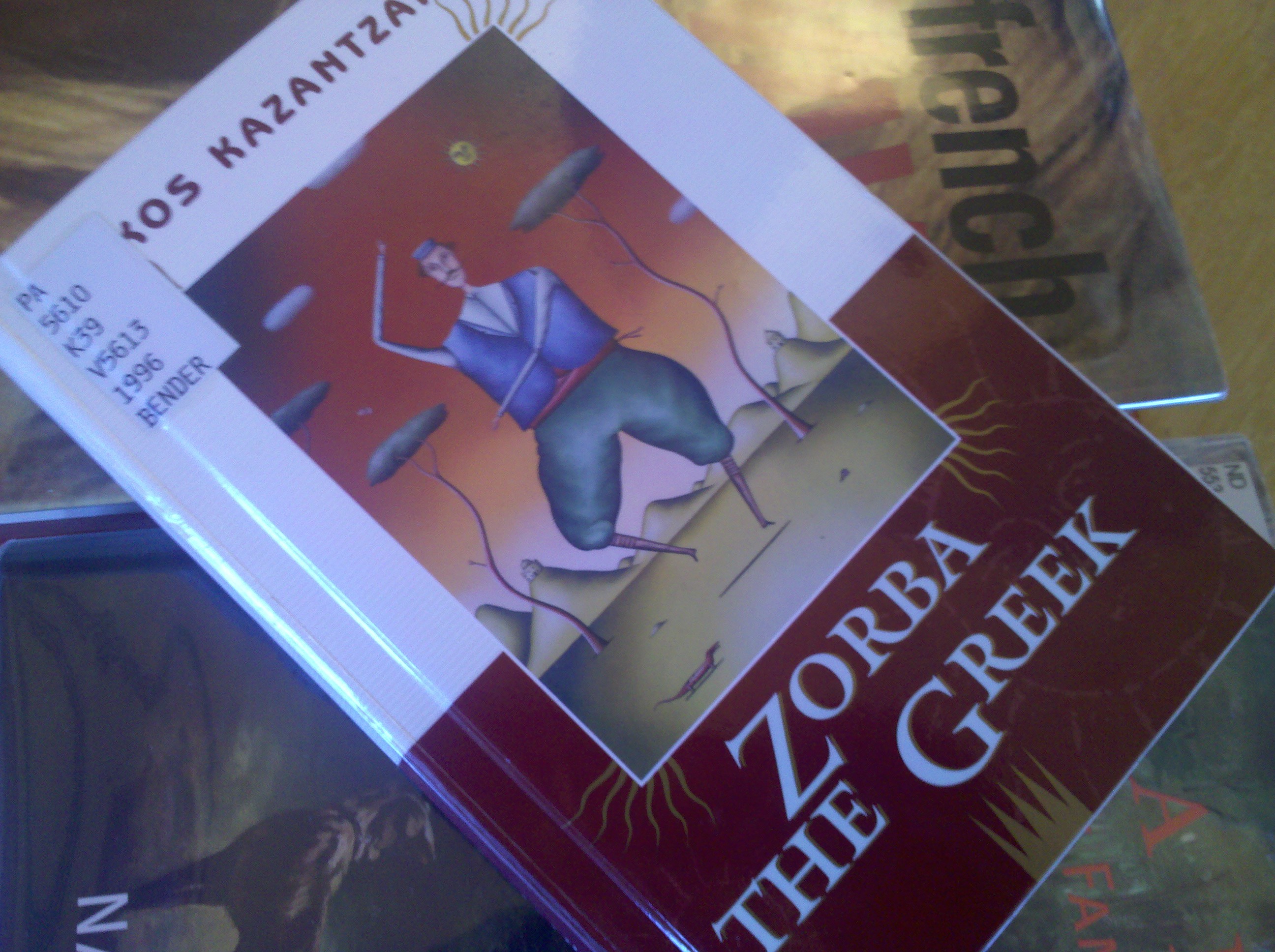} &\includegraphics[width=0.18\linewidth]{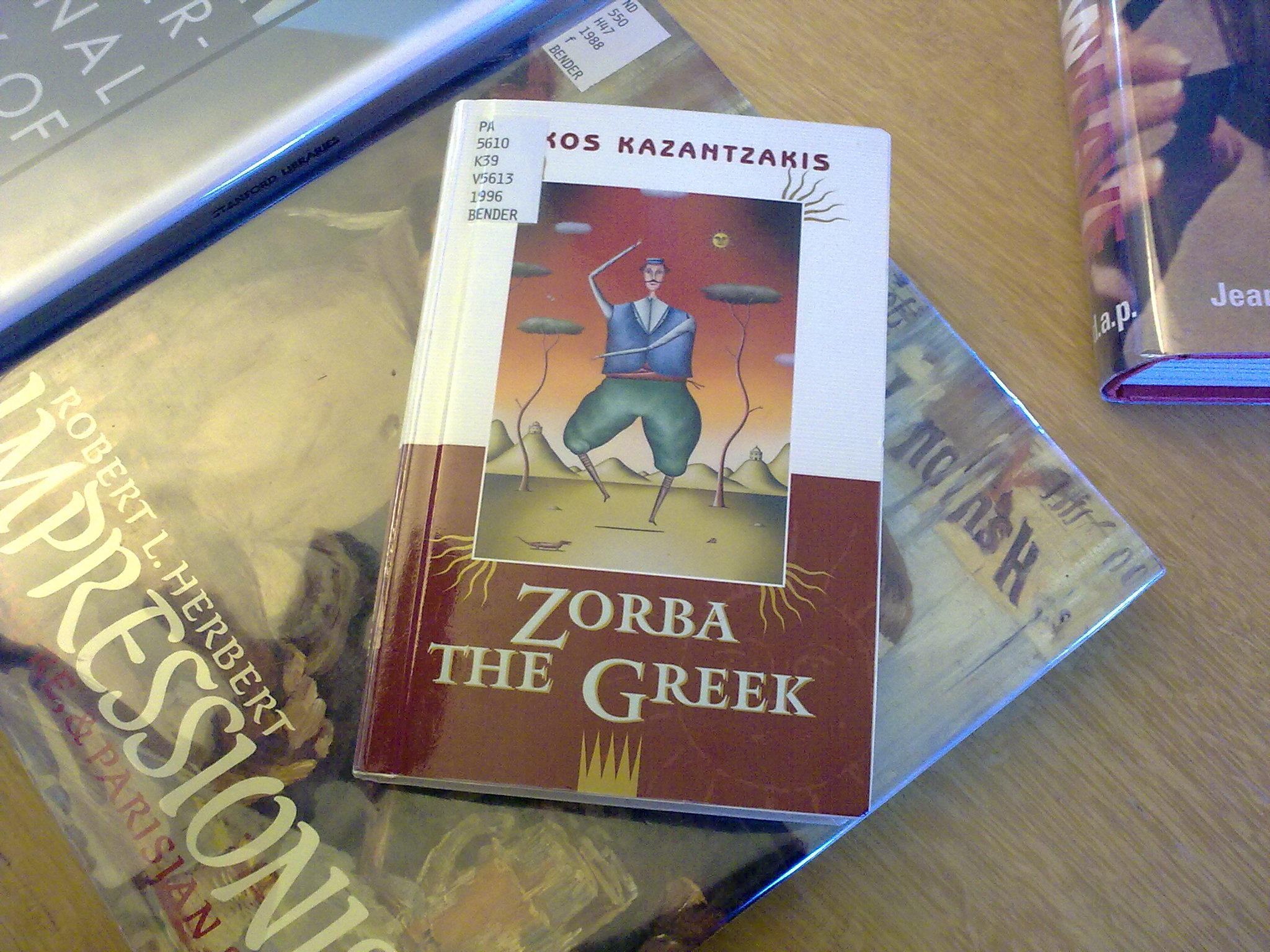}\\

      \includegraphics[width=0.09\linewidth]{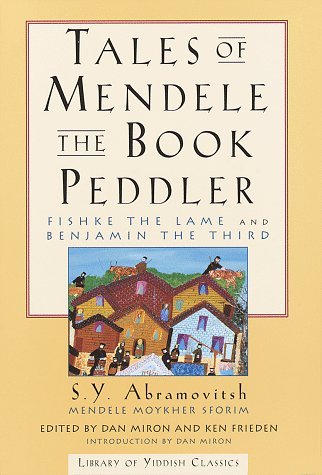} & \includegraphics[width=0.18\linewidth]{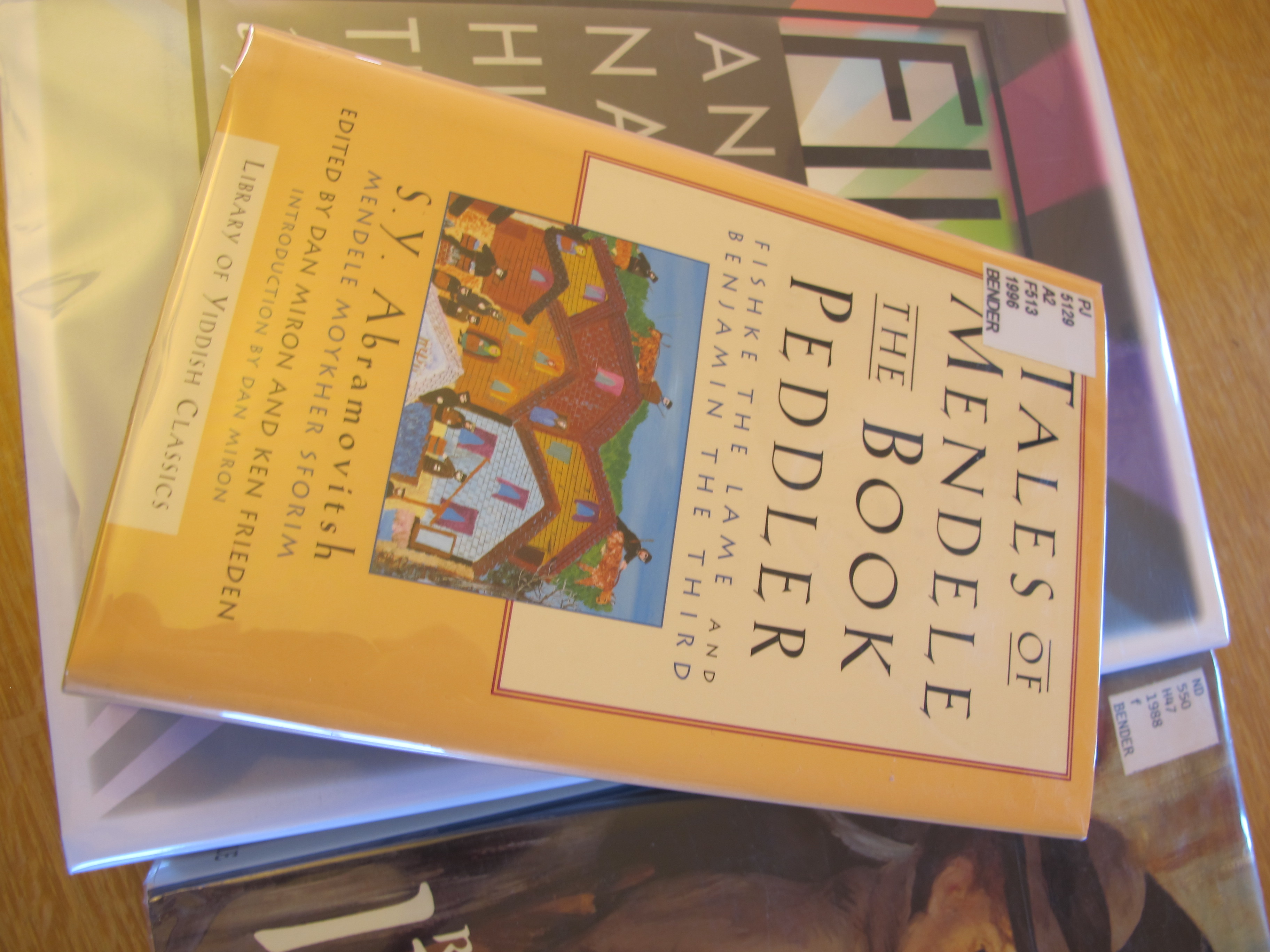} & \includegraphics[width=0.1\linewidth]{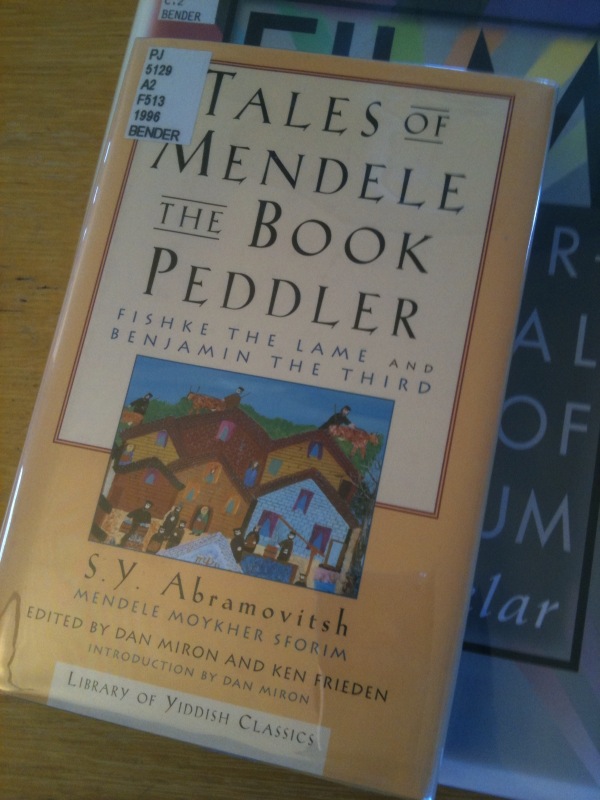} & \includegraphics[width=0.18\linewidth]{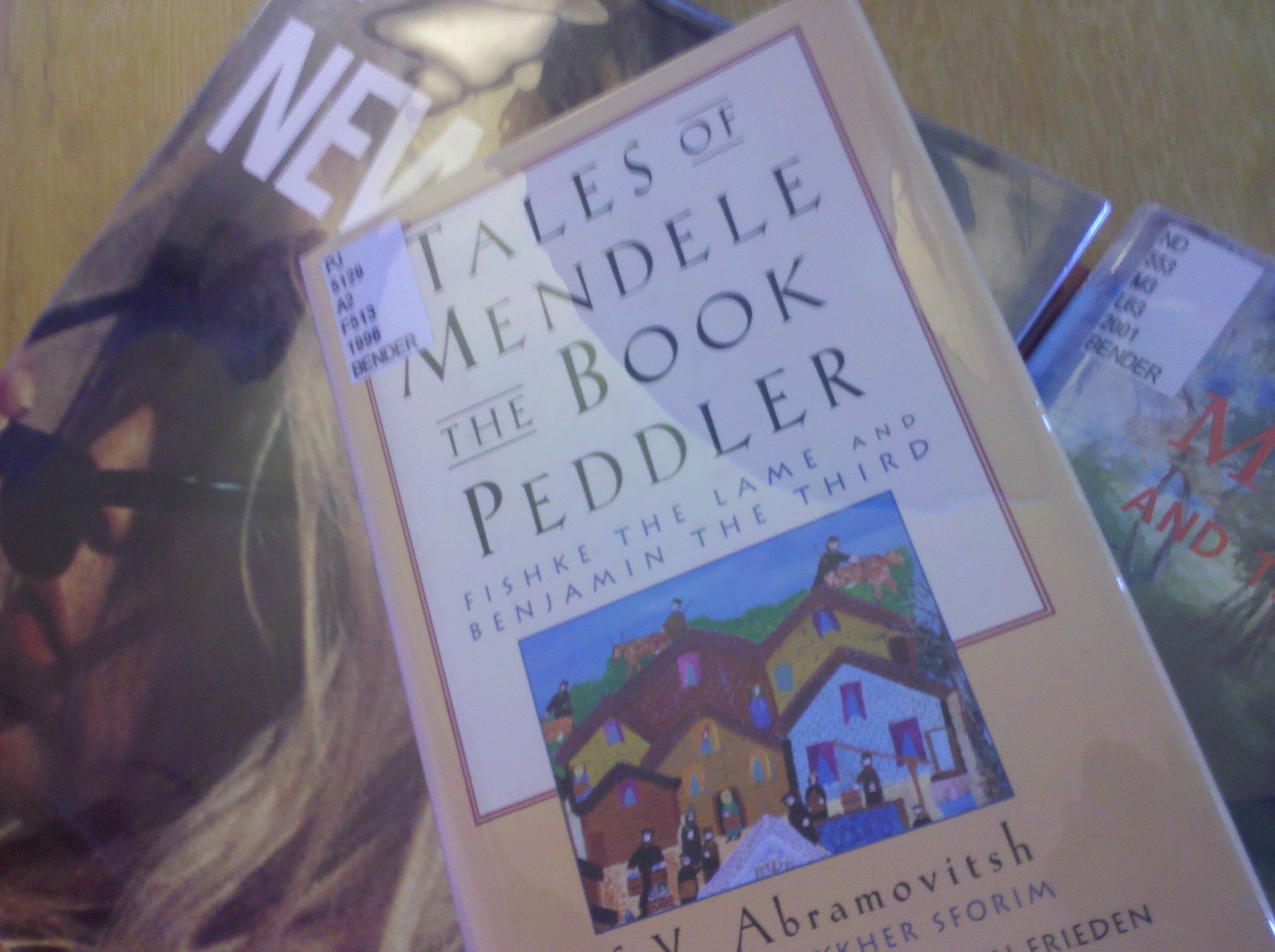} &\includegraphics[width=0.18\linewidth]{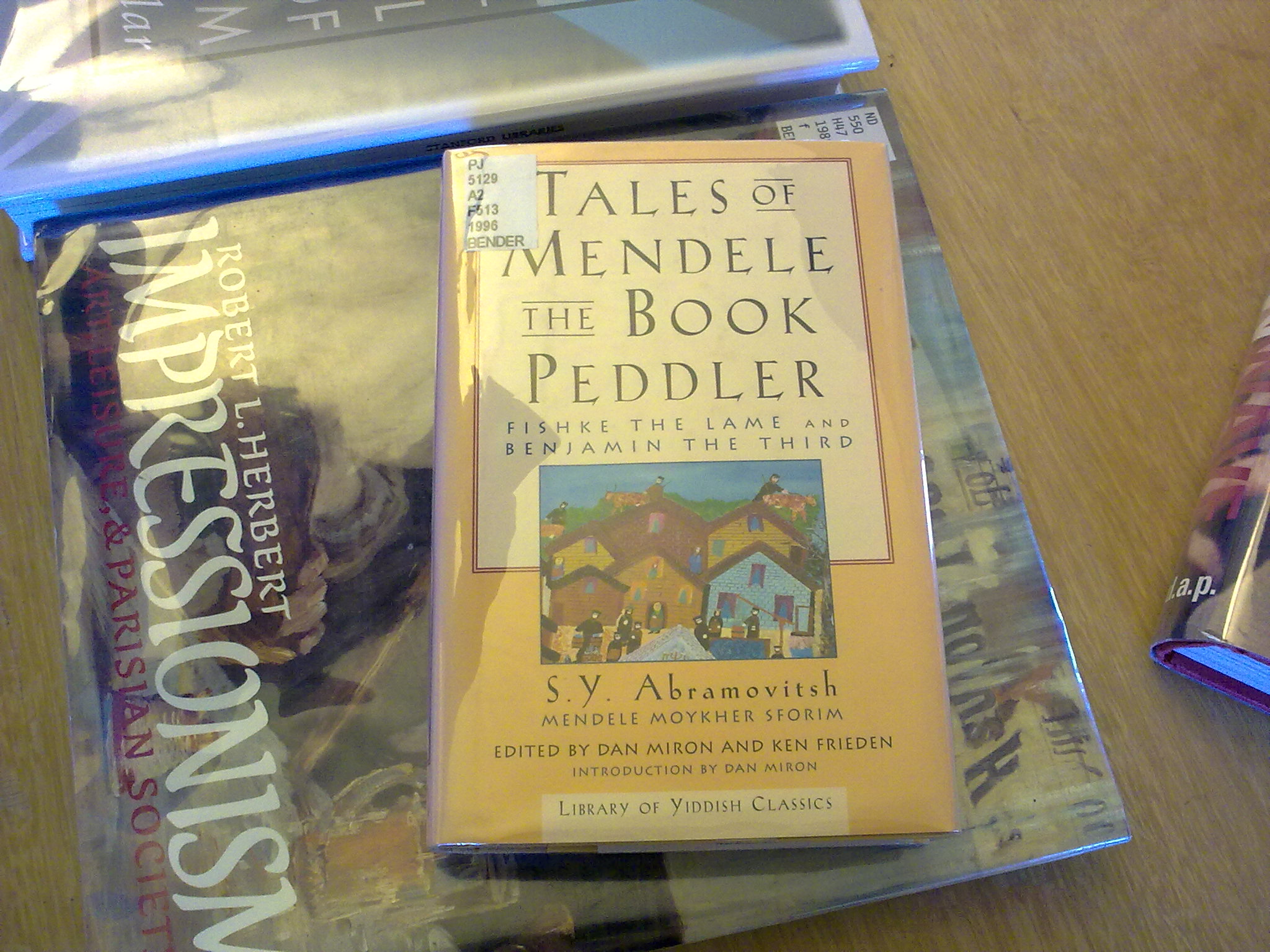}\\

      \includegraphics[width=0.09\linewidth]{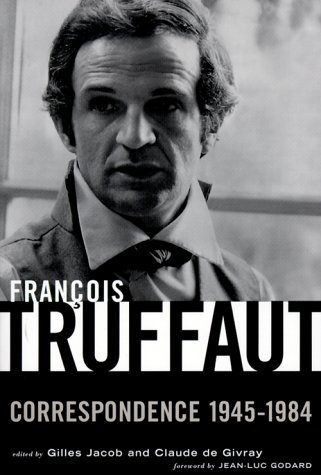} & \includegraphics[width=0.18\linewidth]{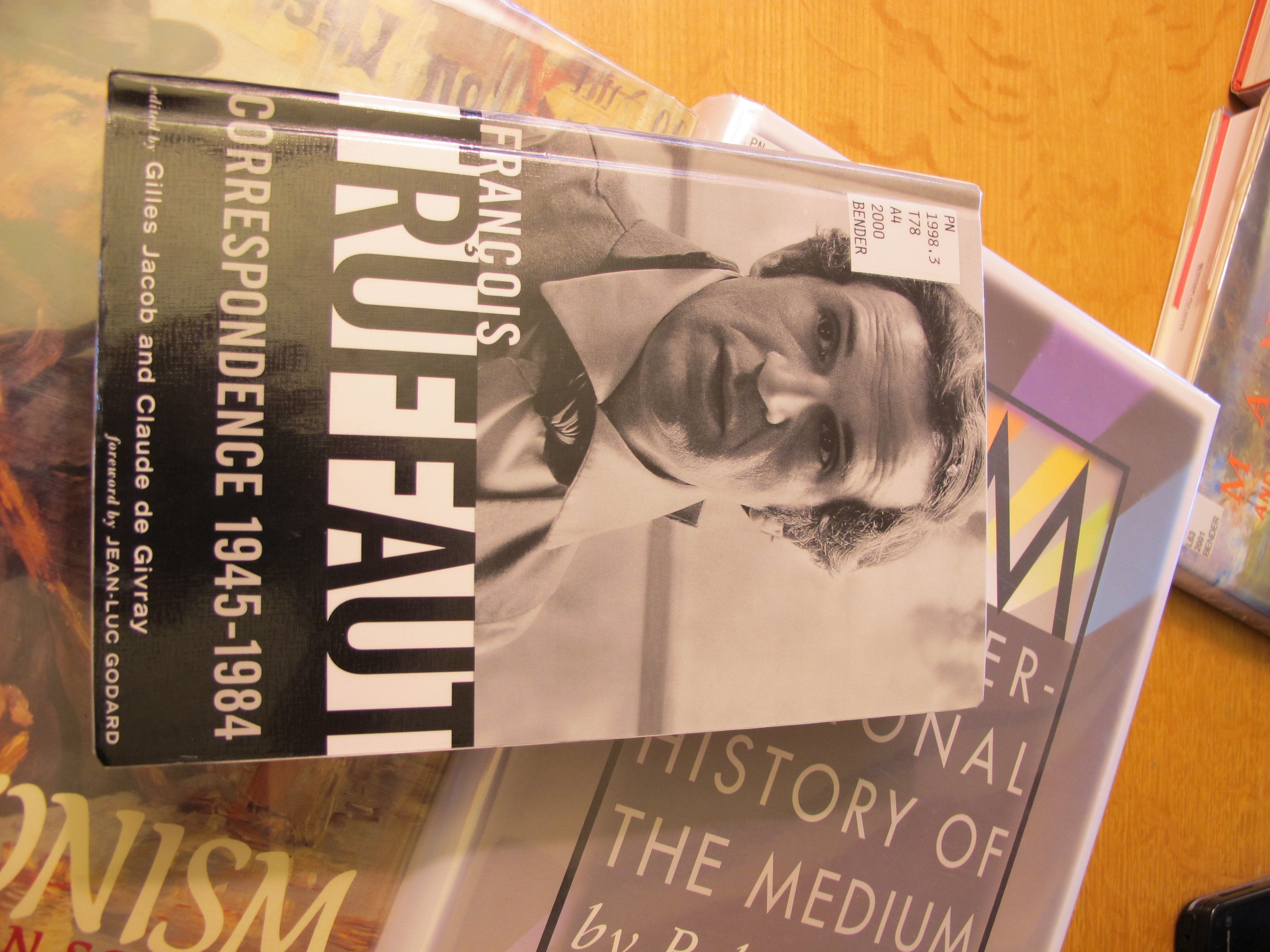} & \includegraphics[width=0.18\linewidth]{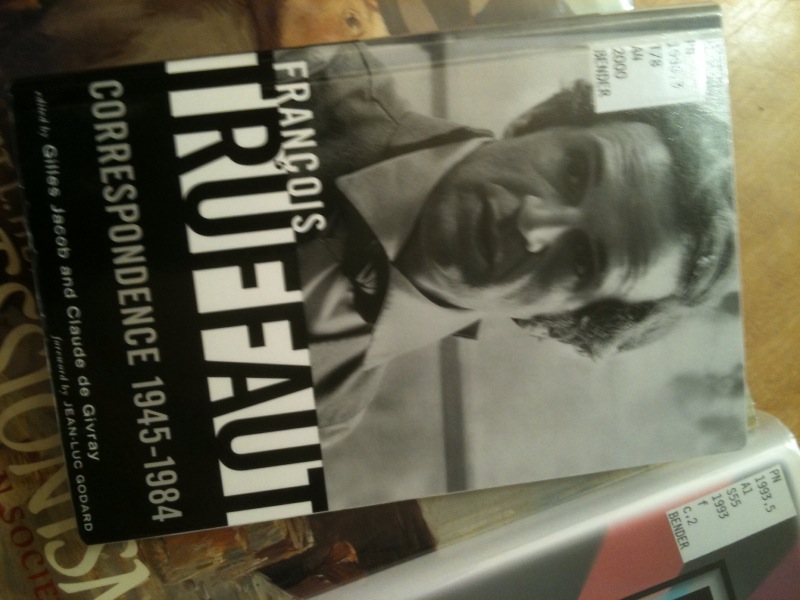} & \includegraphics[width=0.18\linewidth]{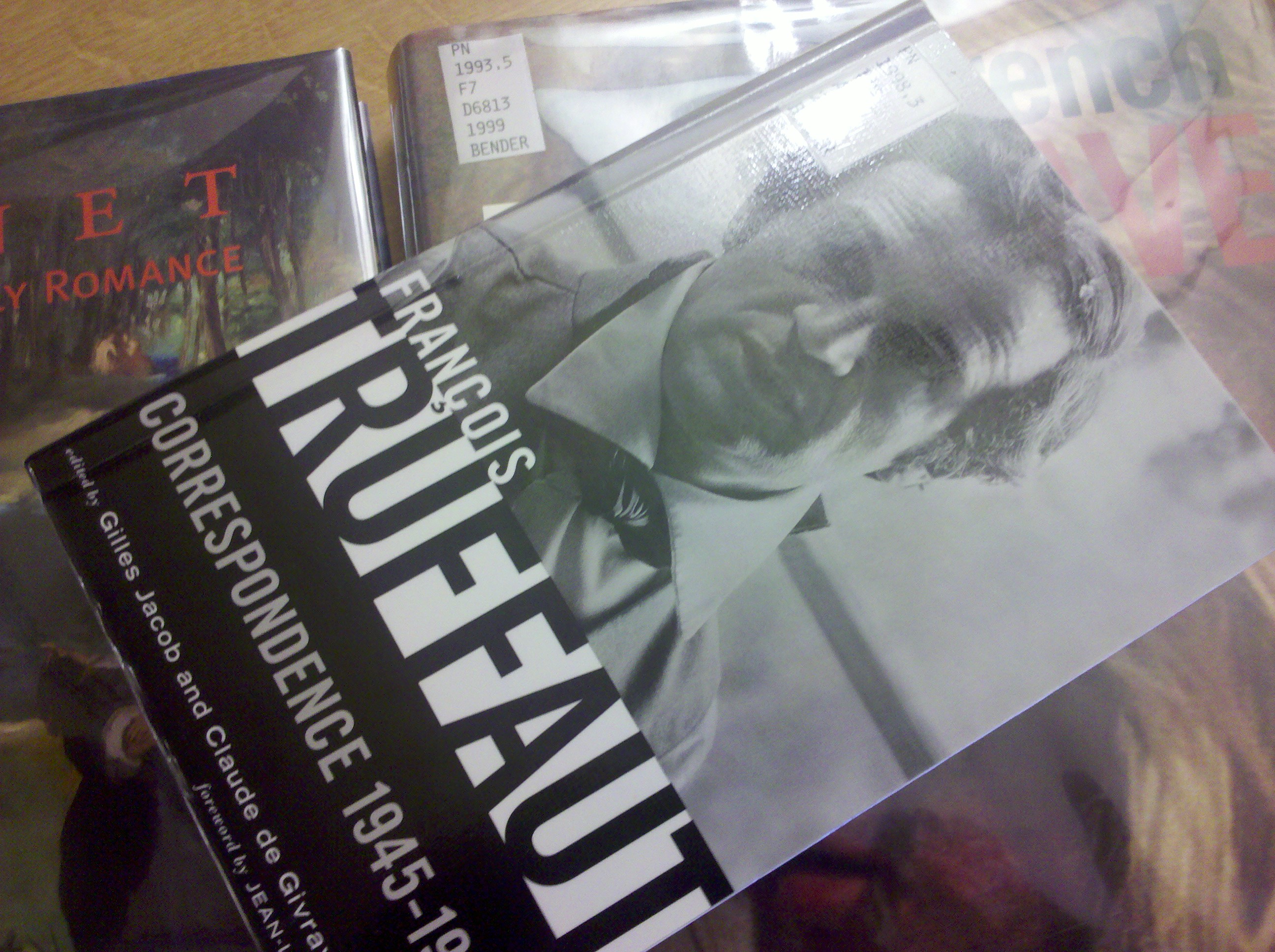} &\includegraphics[width=0.1\linewidth]{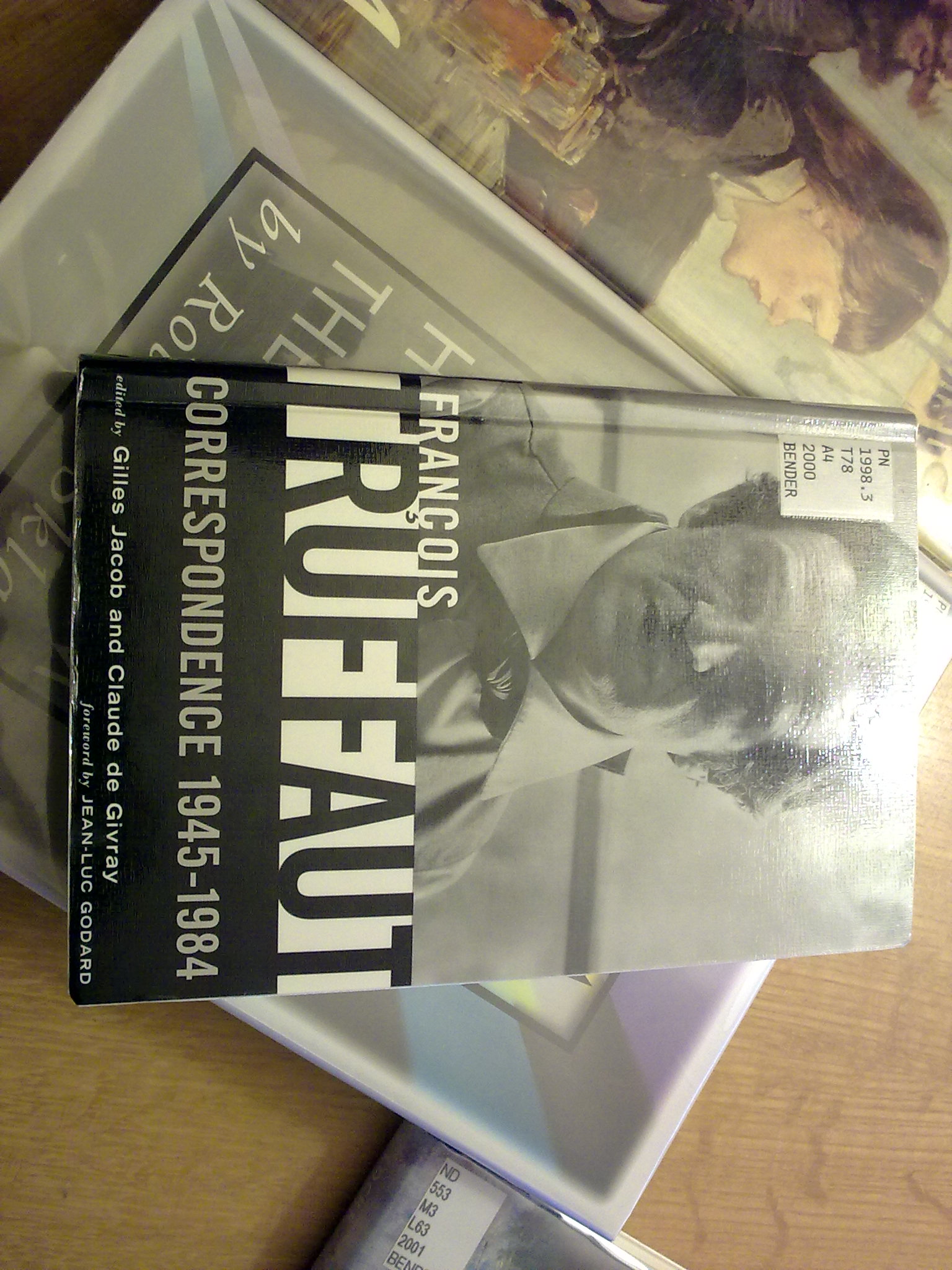}\\

      (a) & (b) & (c) & (d) & (e)
    \end{array}
    $
  \end{center}
  \caption{(a) Reference (b) Canon G11 (c) iPhone 4 (d) Motorola Droid (e) Nokia N5800. Samples from Stanford mobile visual search dataset. Each column corresponds to pictures taken by a specific camera model. Note the orientation, illumination, background variations. Some backgrounds include other books, which makes this a challenging task.}
  \label{figure:StanfordMobileMediaDataset}
\end{figure}

We describe a framework for book cover look-up given a potentially poorly angled, lit, and focused query image.  Example queries can be seen in Fig.~\ref{figure:StanfordMobileMediaDataset}. The query image is a mobile photo of a book taken from various angles and orientations with significant background clutter. Given a query image, we wish to quickly retrieve the corresponding clean catalogue image from a database. 

Our proposed algorithm has two main steps. We first rank our
retrieval results using an ensemble of several BoW features (\ref{sec:initranking},~\ref{sec:reranking}), as these
can be matched quickly and are robust to various distortions. We learn the combining weights using a rankSVM~\cite{Joachims2006RSVM} based formulation. Next, we select a cutoff point in the ranking with
sufficiently high recall and perform more expensive linear-time template
matching on the top results (\ref{subsec:ransacrerank}).

\begin{figure}[t]
  \scriptsize
  \begin{center}
    $
    \begin{array}{cccc}
      \includegraphics[width=0.17\linewidth]{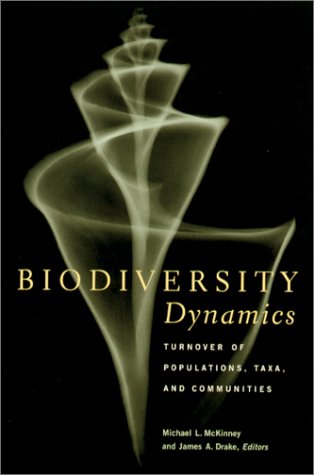} & \includegraphics[width=0.185\linewidth]{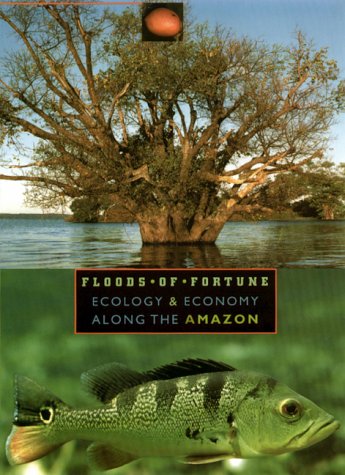} & \includegraphics[width=0.17\linewidth]{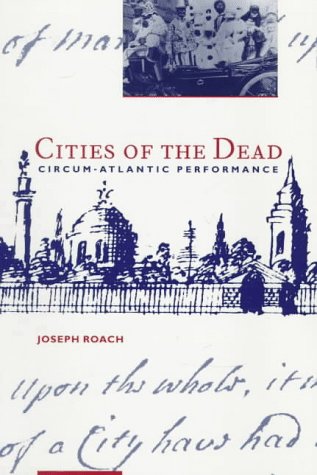} & \includegraphics[width=0.17\linewidth]{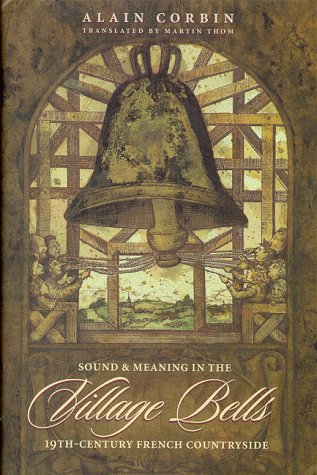} \\

      \includegraphics[width=0.19\linewidth]{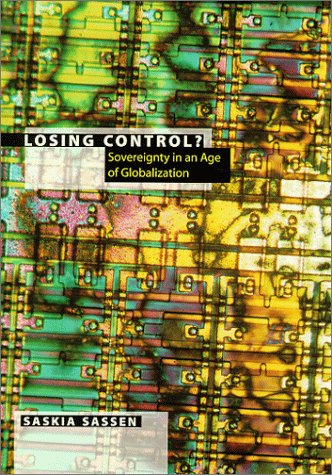} &\includegraphics[width=0.15\linewidth]{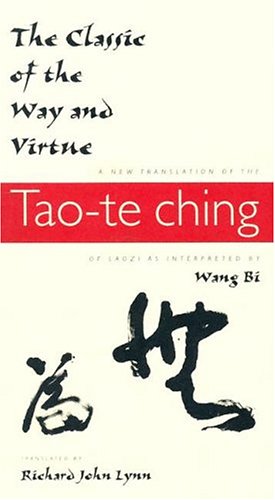}& \includegraphics[width=0.18\linewidth]{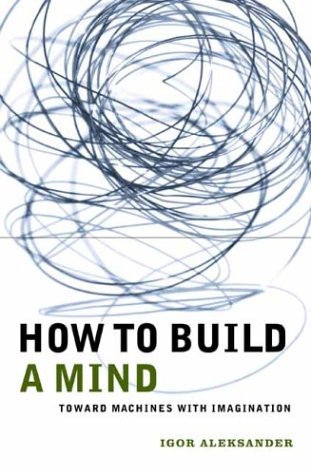} &  \includegraphics[width=0.18\linewidth]{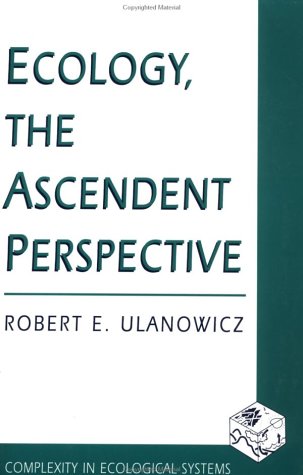}

    \end{array}
    $
  \end{center}
  \caption{Illustration of clean catalogue distractors from openlibrary.org. Using their database, we were able to extract over 100k book covers and their corresponding title and author information.}
  \label{figure:DistractorsSamples}
\end{figure}

\subsection{Dataset}
\label{subsec:dataset}
We apply our method on book cover queries provided by Stanford Mobile
Visual Search dataset~\cite{chandrasekhar2011stanford}. For book covers, this dataset provides up to 4
queries from different mobile devices for 101 book titles, resulting
in a total of 404 queries. To demonstrate the benefit of text-based features in product look-up, we added additional annotations in the
form of title and author names for each book cover. Further, we created a distractor set of 104,132 additional
distractor images taken from openlibrary.org to emulate large databases, each with title and author information (when available). The distractor set was pruned for possible duplicates with the original 101 titles. To our knowledge, this is the first large-scale book cover retrieval dataset to contain author and title text annotation for every single image. The augmented text annotations and distractor images will be released to public for other researchers to test on.

\subsection{Ranking Features}
\label{sec:initranking}
We discuss the BoW features we use to rank catalogue images in a database, given a query. As previously mentioned, we use BoW features as they can be quickly used to generate feature vectors and are compatible with many fast look-up algorithms such as approximate nearest-neighbors. In this work, we focus on the use of VLAD~\cite{jegou2010aggregating} and our proposed method of extracting textual $N$-grams from images. Because the two methods differ significantly in the types of mistakes they make, a hybrid ranking should be able to perform noticeably better than either alone.

\subsubsection{VLAD}

To construct the Vector of Locally Aggregated Descriptors (VLAD)~\cite{jegou2010aggregating}, we first densely extract SURF
descriptors, and cluster them using k-means with $k=256$ to generate the vocabulary. When distractors are used, we use a subset of 15k distractor images to learn the dictionary, otherwise we use only the 101 ground truth catalogue images. The similarity metric is based on the L2 distance between normalized VLAD descriptors.

\begin{figure*}[t]
  \centering
  \includegraphics[width=\textwidth]{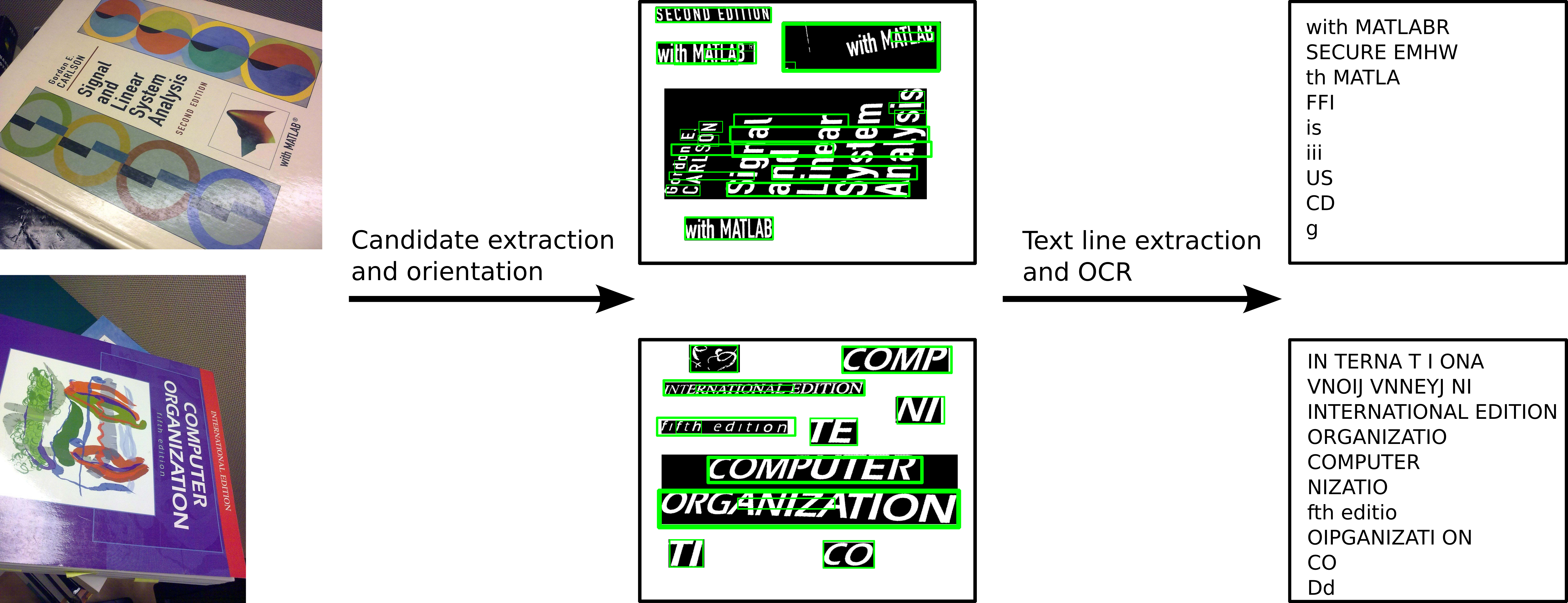}
  \caption{Two examples of text being extracted from adversarial mobile queries. Given a query image, blocks of text are first identified and oriented using a radon-transform based heuristic. Then we identify roughly co-linear characters as lines and feed them through Tesseract OCR to obtain text output. The entire set of extracted text from the image will be treated as a text document in a standard document look-up problem to match against strings of author and book title annotations. The shown text and detection outputs are only a subset of the actual output on the presented images. }
  \label{fig:text_extraction_pipeline}
\end{figure*}
\subsubsection{OCR}

\boldhead{Localizing Oriented Text}
Performing OCR on non-cooperative queries is especially
challenging -- lines of text can be oriented any which way, appear in
any location, and vary
significantly in font size and style within the same image. Further,
specularities and distortions can block out important characters and text
appearing in the background can contribute significant noise. Here, we
propose a robust pipeline for matching text information given such
adversarial conditions.

We first use ~\cite{gomez2013multi} to extract clusters of text in the form of binary masks. Unlike most available
text localization methods, this extracts text in an
orientation-agnostic manner rather than assuming horizontal
alignment.
 
Next, we need to re-orient the text so that we can extract independent lines. We orient the text clusters to horizontal alignment via projection
analysis. We compute a radon transform and select the angle of the line
with the least projected area. The assumption behind this is
that lines of text will be longer than they are tall. This also works
for clusters with multiple lines, but occasionally fails for
tall and slim blocks of text. One such failed case can be seen in Fig.~\ref{fig:text_extraction_pipeline}.

To extract individual lines of text, we use simple clustering of proximal characters. Our
goal is to achieve as high recall as possible, without regard to
overalapping duplicates and false positives. Initially, we identify MSER~\cite{MSER} regions as potential characters within the segmented
image. To group character candidates into lines, we
attempt to combine regions of similar height 
if they are adjacent or if their bases have a close $y$ value. We rule out
unrealistic line candidates based on aspect ratio (length/width $>$ 15).

Finally, we crop out the lines of text and feed them through
Tesseract OCR~\cite{smith2007overview} to extract text. Because the oriented text lines could be upside down, we also feed in the 180 degree rotated version of each cropped line to obtain a separate output. We find that Tesseract works best with pre-localized text lines as its
internal text localization isn't robust enough to handle unoriented input.

\boldhead{Extracting Tokens}
Because we expect the OCR output to be very noisy and corrupted, we propose to match character
N-grams as opposed to entire words.  Character N-grams are commonly used as a low-computation and low memory solution to approximate string-matching~\cite{Navarro00indexingmethods}.  To do this, we run a sliding window of size $N$ across each word with sufficient length and ignore non-alphabets. For a 3-gram example, the phrase, ``I like turtles'' would be broken down into ``lik,'' ``ike,'' ``tur,'' ``urt,'' ``rtl,'' ``tle,'' ``les.''
The benefit of this method is that we can still achieve a match as
long as OCR is able to correctly return a sequence of at least 3
characters in a row. For simplicity, we also ignore case by converting everything to lowercase.

\boldhead{Efficient Matching}
In our dataset, we assume each book cover to be annotated with book title and author information. This information allows us to use a text-based document retrieval approach.

As is common in document retrieval, we retrieve our documents by
taking the inner product of tf-idf~\cite{tfidf} weighted normalized histogram of
N-grams. This normalized inner product can be computed very efficiently using
inverted file indexing, and can be trivially converted to a 
normalized euclidean distance.

Let $\vec{f}$ be the un-normalized histogram of $N$-grams for each
document. We use the following scheme to compute a normalized
similarity score between query and document:
\begin{center}
\begin{tabular}{ l | l }
Query & Document \\
\hline
$N_2(N_1(\vec{f})^T\vec{\gamma})$ & $N_2(N_2(\vec{f})^T\vec{\gamma})$ \\
\end{tabular}
\end{center}
where $N_1$ and $N_2$ are functions for computing L1 and L2
normalization respectively. The vector $\vec{\gamma}$ is the vector of 
idf-weights. For each unique $N$-gram $g$, we compute its corresponding
idf weight as $\vec{\gamma}(g) = \ln{\frac{|D|}{|d \in D : g \in d| }}$ : the natural log
of the number of documents in the database divided by the number
containing the $N$-gram $g$. We found that the choice of pre-idf
weighting normalization matters little, but that the final
normalization should be L2 as we are using euclidean distances.

\boldhead{Robustness to False Positives}
Due to the nature of our inner-product based matching mechanism, we are able to have our OCR pipeline focus more on recall and to disregard the existence of the false positives. As can be seen in some of the example output in Fig.~\ref{fig:text_extraction_pipeline}, many false positives will appear as either single characters or improbable character sequences due to accidental recognition on background patches or our consideration of 180 degree rotated text lines. The improbable sequences will have little effect on the matching as said sequences will rarely occur in the retrieval set. The single character false positives will be completely ignored by setting a sufficiently large $N$-gram length.

\boldhead{Runtime}
While our code is not fully optimized for speed, the main components of our OCR pipeline are fairly efficient and very easily parallelizable. This aspect is important for commercial search systems where query times should not exceed a couple of seconds. Our main bottleneck is the extraction of multiple text candidates from each query image. While fairly clean images will finish in 2-3 seconds, images with many blob-like regions will trigger large amounts of false positives, each of which will need to be independently run through Tesseract and may take up to 30 seconds total in our unoptimized setup. Further, while it is clear that independent calls to Tesseract can be run in parallel, which we have implemented, the runtime will then be lower-bounded by that of a single call to Tesseract.

The matching component is implemented with an inverted file index and, when properly implemented with hash tables, will compute all inner products with all 101 book titles in less than a second per query.

\begin{figure*}[t]
  \centering
  \includegraphics[width=\textwidth]{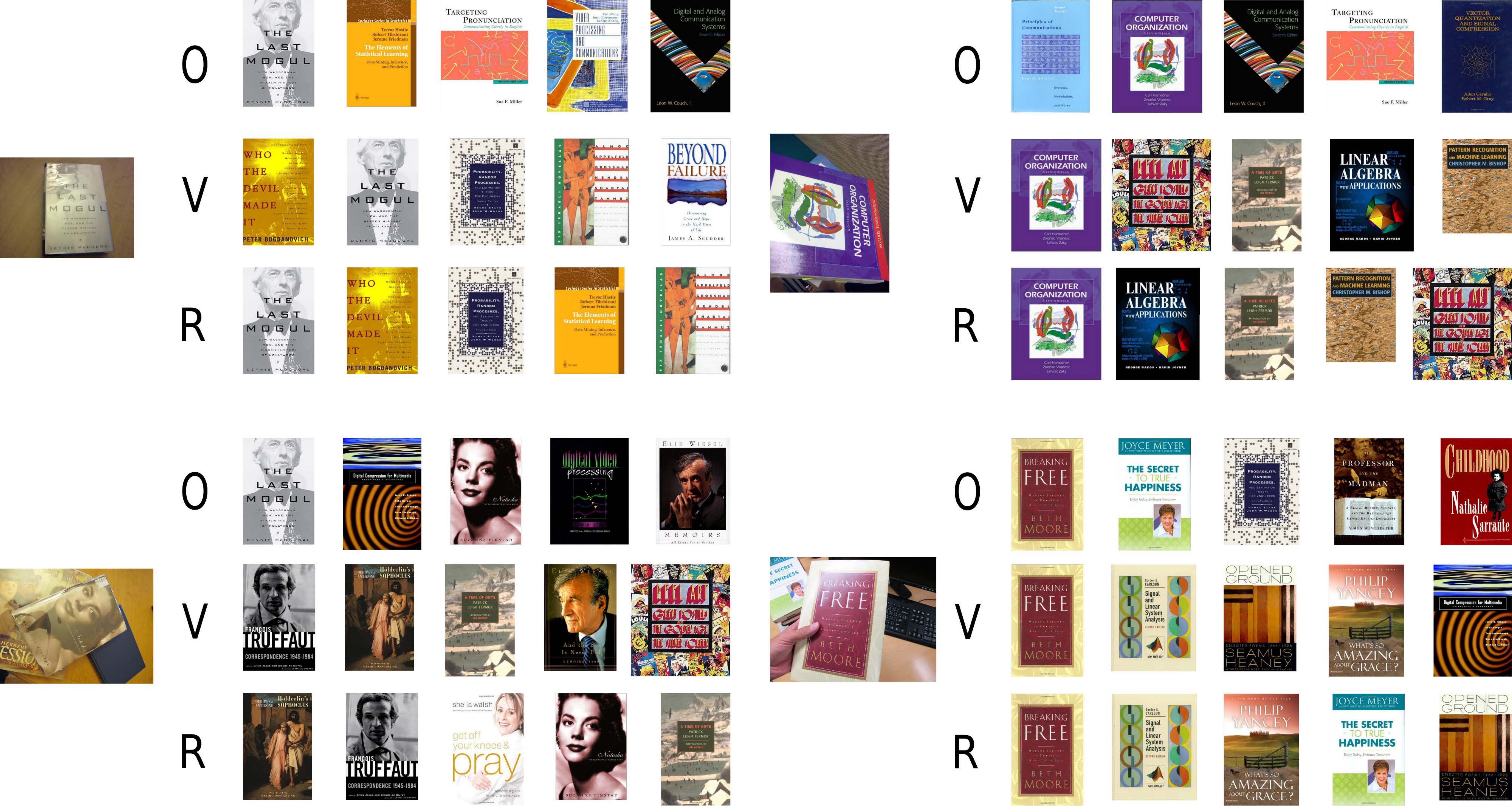}
  \caption{Top 5 queried results on from the dataset for each feature type in descending rank from left to right using the mobile image to the left of each 'V'. Rows starting with 'O' refer to top results ranked only with OCR-3, 'V' for only VLAD, and 'R' for rankSVM re-ranking. Note the different types of mistakes made by OCR and VLAD. The former makes mistakes based on similar words appearing in a book. The latter makes more visual pattern based mistakes.}
  \label{fig:top_examples}
\end{figure*}

\subsection{Re-Ranking}
\label{sec:reranking}
We define the vector: \[\Phi(x,y) = [  S_1(x,y)~S_2(x,y)~\cdots~S_M(x,y) ]^T \]
where each $S(x,y)$ represents a similarity measure from any feature type
such as VLAD or OCR between query image $x$ and reference image $y$. We wish to learn an optimal weighting for
$\Phi(x,y)$ for the purposes of ranking and retrieval. 

While the weighting can be easily determined through trial-and-error
for two or three distance measures, we formulate the problem as a more
scalable learning problem. Specifically we optimize the following objective:
\begin{equation}
  \begin{aligned}
    &\argmin_{\vec{w}} & &\frac{1}{2}\vec{w}^T \vec{w} + C\sum{\xi_{i,j}} \\
    &\text{s.t.}\\
    &&&\vec{w}^T(\Phi(x_i, y_i)-\Phi(x_i, y_j)) > 1-\xi_{i,j}~\forall i, j \neq i\\
    &&&\xi_{i,j} \geq 0~\forall i,j
  \end{aligned}
\end{equation}

Simply put, we wish to learn the optimal weighting $\vec{w}$ for our
combined distance metric $\Phi(x,y)$ such that the similarity between a
correct query/reference match ($i=j$) is always greater than that of an
incorrect one ($i\neq j$). This is very similar to the objective for
structured-SVMs and can be efficiently optimized using the SVMRANK
package~\cite{Joachims2006RSVM}. This model is referred to as the rankSVM model in later sections.

Because our dataset does not have a train/test split, we compute our
final results using a simple two fold cross validation. The query set of mobile book images is split into two parts and we alternate their roles by training on one and evaluating on the other. This ensures that no query image was evaluated by a model trained on itself.

\subsubsection{RANSAC rectification with HOG Matching}
\label{subsec:ransacrerank}
We finalize our pipeline with a brute-force RANSAC instance matching
and HOG template matching based search on the top $K$ results. If the BoW approach is sufficient to
bring the correct result within the top $K$, then it may be worth it to incur a constant cost to refine the ranking using a more expensive method. To match the HOG
representation, we first resize the reference and rectified query to
256 by 256 pixels, and compute the inner product of normalized HOG
representations with 8 orientations, 8 by 8 pixels per cell, and 2 by
2 cells per block.

\section{Results}
In the following section, we test performances of individual components of our method on our data, first without distractors, then with them in subsection~\ref{subsec:distractor_perf}. In our result figures, we look at how large a retrieval set needs to be grown to obtain the correct catalogue image corresponding to the mobile query image. To do this, we plot the fraction of queries that were correctly matched against the size of the retrieval set. In Table~\ref{tbl:detailed_pnts}, we detail the exact retrieval rate of various methods at sizes 1, 5, 10, and 20. Example retrievals for select queries can be seen in Fig.~\ref{fig:top_examples}.

\subsection{OCR $N$-gram size}
\begin{figure}[t]
  \begin{center}
    \begin{tikzpicture}
    \footnotesize
      \pgfplotsset{small}
      \pgfplotstableread{data/ocr_ranks.dat}
      \ocrranktable
      \begin{axis}[title=OCR $N$-gram Retrieval Performance,
        xlabel=Number of Top Retrieved,
        ylabel=Correct Retrieval Rate,
        ylabel near ticks,
        legend columns = 1,
        legend entries={$N=1$,$N=2$, $N=3$, $N=4$, $N=5$},
        legend pos= outer north east,
                legend style={font = \small}]

        \addplot+[mark=none, smooth] table[x = retrieved,, y = ocr-1] from \ocrranktable;
        \addplot+[mark=none, smooth] table[x = retrieved,, y = ocr-2] from \ocrranktable;
        \addplot+[mark=none, dashed] table[x = retrieved,, y = ocr-3] from \ocrranktable;
        \addplot+[mark=none, dash pattern=on 3pt off 6pt on 6pt off 6pt] table[x = retrieved,, y = ocr-4] from \ocrranktable;
        \addplot+[mark=none, dash pattern=on 1pt off 3pt on 3pt off 3pt] table[x = retrieved,, y = ocr-5] from \ocrranktable;
      \end{axis}
    \end{tikzpicture}
  \end{center}
\caption{Retrieval performance of $N$-gram character OCR at different $N$}
\label{fig:OCR-N-comp}
\end{figure}
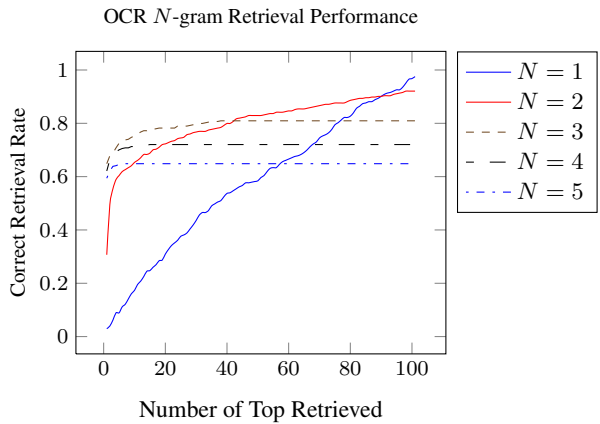

We test the performance of OCR for the $N$-gram character representation
using varying sizes of $N$ in Fig.~\ref{fig:OCR-N-comp}. We find $N=3$ to
perform best for our purposes on the annotated books dataset. Little
to no additional gain in performance was observed at larger sizes $N$,
while recall declined as expected. $N=2$ performed reasonably with
better recall than $N=3$, but accuracy at top 1 retrievals dropped from
.65 to .31. $N=1$ was mostly noise as it had few means of dealing with the noise from our text-extraction process. However, it is worth noting that by not pruning out anything, $N=1$ achieves near-perfect recall at the end. We use $N=3$ (ocr-3) in all future experiments unless otherwise noted.

\subsection{VLAD+OCR}
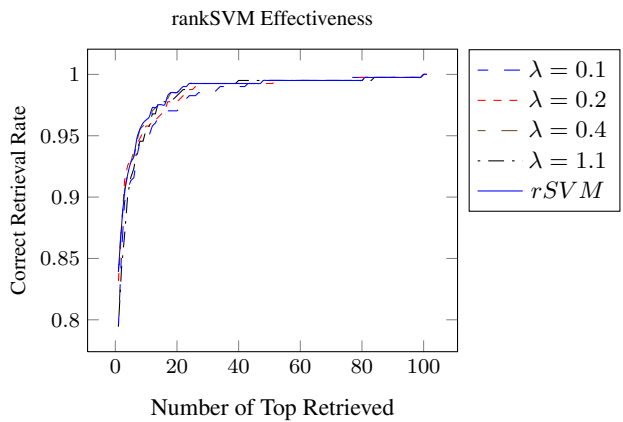
\begin{figure}
  \begin{center}
    \begin{tikzpicture}
    \footnotesize
      \pgfplotsset{small}
      \pgfplotstableread{data/hybrid_ranks.dat}
      \hybridranktable
      \begin{axis}[title=rankSVM Effectiveness,
        xlabel=Number of Top Retrieved,
        ylabel=Correct Retrieval Rate,
        ylabel near ticks,
        legend columns = 1,
        legend entries={$\lambda=0.1$, $\lambda=0.2$, $\lambda=0.4$,
          $\lambda=1.1$, $rSVM$},
        legend style={font = \small},
        legend pos= outer north east]
        \addplot+[mark=none, dash pattern=on 4pt off 6pt on 6pt off 6pt] table[x = retrieved,, y = lamb_01] from \hybridranktable;
        \addplot+[ mark=none, dashed] table[x = retrieved,, y = lamb_02] from \hybridranktable;
        \addplot+[ mark=none, dash pattern=on 3pt off 6pt on 6pt off 6pt] table[x = retrieved,, y = lamb_04] from \hybridranktable;
        \addplot+[ mark=none, dash pattern=on 1pt off 3pt on 6pt off 3pt] table[x = retrieved,, y = lamb_11] from \hybridranktable;
        \addplot+[ mark=none, smooth] table[x = retrieved,, y = rsvm] from \hybridranktable;
      \end{axis}
    \end{tikzpicture}
  \end{center}
\caption{Retrieval performance of combined OCR and VLAD rankings comparing the learned rankSVM model against models with hard-coded $\lambda$ values.}
\label{fig:OCR-VLAD-hybrid-rsvm}
\end{figure}

First, we look at the results of combining OCR and VLAD scores with our learned weights
in Fig.~\ref{fig:OCR-VLAD-hybrid-rsvm}. Here, we compare our learned model with the model resulting from several values of a hard-coded $\lambda$. The final scores of the hard-coded models are $S_{\text{VLAD}}+\lambda S_{\text{OCR}}$. Our learned model results were generated with two-fold cross validation and no distractors were used in this experiment. Results show that the learned model performs comparably to the best-performing hard-coded models.

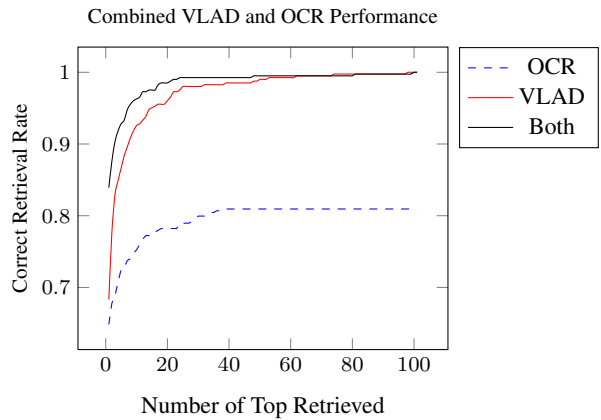
\begin{figure}
  \begin{center}
    \begin{tikzpicture}
    \footnotesize
      \pgfplotsset{small}
      \pgfplotstableread{data/ocr_ranks.dat}
      \ocrranktable
      \pgfplotstableread{data/vlad_ranks.dat}
      \vladranktable
      \pgfplotstableread{data/hybrid_ranks.dat}
      \hybridranktable
      \begin{axis}[title=Combined VLAD and OCR Performance,
        xlabel=Number of Top Retrieved,
        ylabel=Correct Retrieval Rate,
        ylabel near ticks,
        legend columns = 1,
        legend entries={OCR, VLAD, Both },
                legend style={font = \small},
        legend pos= outer north east]
        \addplot+[mark=none, dashed] table[x = retrieved,, y = ocr-3] from \ocrranktable;
        \addplot+[mark=none, smooth] table[x = retrieved,, y = VLAD] from \vladranktable;
        \addplot+[color=black, mark=none, smooth] table[x = retrieved,, y = rsvm] from \hybridranktable;
      \end{axis}
    \end{tikzpicture}
  \end{center}
\caption{Retrieval performance of combined OCR and VLAD rankings}
\label{fig:OCR-VLAD-hybrid}
\end{figure}

Next, we compare the performance of a combined OCR and VLAD rankSVM model to that of its individual features in Fig.~\ref{fig:OCR-VLAD-hybrid}. The combined model performs at least as well as the individual components, with a significant improvement in the top-1 retrieval result. In Table~\ref{tbl:detailed_pnts}, we see that the combined model (rankSVM) has a retrieval rate of 0.84 using only the top retrieval, as compared to 0.65 and 0.68 for ocr-3 and VLAD respectively. 

We visualize the top queries of each ranking feature individually in Fig.~\ref{fig:top_examples} to try to understand the improvement. As can be seen, OCR and VLAD make very different types of errors. Most OCR-based errors involve assigning high scores to other books with similar-sounding titles. VLAD on the other hand tends to assign higher scores to candidates with similar visual patterns. Often, the only candidate they agree on is the correct one, a pattern that is exploited by the rankSVM combined model.


\subsection{Fine-Grained Reranking with RANSAC and HOG}
\begin{figure}
  \begin{center}
    \begin{tikzpicture}
    \footnotesize
      \pgfplotsset{small}
      \pgfplotstableread{data/hybrid_ranks.dat}
      \hybridranktable
      \pgfplotstableread{data/rr_rank.dat}
      \rrranktable
      \begin{axis}[title=RANSAC Reranking of Top K,
        xlabel=Number of Top Retrieved,
        ylabel=Correct Retrieval Rate,
        ylabel near ticks,
        legend columns = 1,
        legend style={font = \small},
        legend entries={OCR+VLAD, $K=5$, $K=10$, $K=15$, $K=101$},
        legend pos= south east ]
        \addplot+[ color = black,mark=none, dashed] table[x = retrieved,, y = rsvm]
        from \hybridranktable;
        \addplot+[ mark=none, smooth] table[x = retrieved,, y = rr-5]
        from \rrranktable;
        \addplot+[ color=magenta,mark=none, dash pattern=on 3pt off 6pt on 6pt off 6pt] table[x = retrieved,, y = rr-10]
        from \rrranktable;
        \addplot+[ mark=none, dash pattern=on 1pt off 3pt on 6pt off
        3pt] table[x = retrieved,, y = rr-15] from \rrranktable;
        \addplot+[ mark=none, smooth] table[x = retrieved,, y = rr-101] from \rrranktable;
      \end{axis}
    \end{tikzpicture}
  \end{center}
\caption{Retrieval performance of RANSAC reranking}
\label{fig:Ransac-rr}
\end{figure}
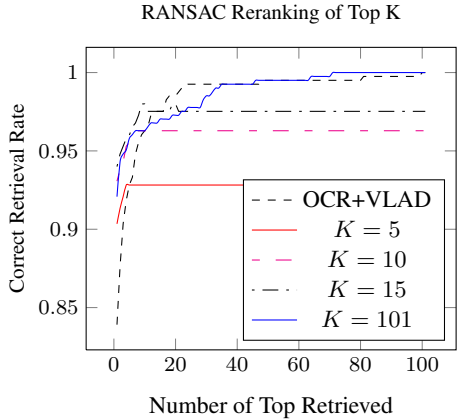

We look at the effects of our previously described RANSAC+HOG reranking in
Fig.~\ref{fig:Ransac-rr}. As this method is time consuming compared to BoW approaches, we use it to rerank the only the top $K$=5, 10, and 15 results. In most cases, if the ground truth falls within the top $K$ retrievals, it will be re-ranked to the top if it isn't already assuming the RANSAC localization was successful. We also compute an expensive upper
bound by re-ranking over all 101 catalogue images. 

As can be seen in Fig.~\ref{fig:Ransac-rr}, the reranked result curves will flatten out due to the fixed max window size $K$. However, often it will flatten out even before reaching $K$ retrievals. This is because the method does not have a 100\% success rate even when the ground truth is included in the top $K$. This can be easily seen in the curve for $K=101$ in how it performs worse than that of the smaller $K$ values. 

The problem with this method can be explained by its two failure points. First, if RANSAC fails to localize, then it will be impossible for HOG to find a match. Next, because we are using raw unweighted HOG features instead of a set of trained HOG weights, the model will tend to assign high scores to highly textured candidates. While there is little that can be done about the shortcomings of RANSAC, it is likely that replacing the HOG filters with an exemplar detector such as with~\cite{malisiewicz2011ensemble} or ~\cite{hariharan2012discriminative} could rectify the latter issue.

\begin{table*}
\centering
\footnotesize
\pgfplotstableread{data/ocr_ranks.dat}
\ocrranktable
\pgfplotstableread{data/vlad_ranks.dat}
\vladranktable
\pgfplotstableread{data/hybrid_ranks.dat}
\hybridranktable
\pgfplotstableread{data/rr_rank.dat}
\rrranktable
\pgfkeys{/pgf/number format/.cd,fixed,precision=2}
\pgfplotstablecreatecol[copy column from
table={\ocrranktable}{ocr-1}]{ocr-1}{\rrranktable}
\pgfplotstablecreatecol[copy column from
table={\ocrranktable}{ocr-2}]{ocr-2}{\rrranktable}
\pgfplotstablecreatecol[copy column from
table={\ocrranktable}{ocr-3}]{ocr-3}{\rrranktable}
\pgfplotstablecreatecol[copy column from
table={\ocrranktable}{ocr-4}]{ocr-4}{\rrranktable}
\pgfplotstablecreatecol[copy column from table={\ocrranktable}{ocr-5}]{ocr-5}{\rrranktable}
\pgfplotstablecreatecol[copy column from table={\ocrranktable}{ocr-3-d}]{ocr-3-d}{\rrranktable}
\pgfplotstablecreatecol[copy column from table={\hybridranktable}{rsvm}]{rsvm}{\rrranktable}
\pgfplotstablecreatecol[copy column from table={\hybridranktable}{rsvm-d}]{rsvm-d}{\rrranktable}
\pgfplotstablecreatecol[copy column from table={\vladranktable}{VLAD}]{VLAD}{\rrranktable}
\pgfplotstablecreatecol[copy column from table={\vladranktable}{VLAD-d}]{VLAD-d}{\rrranktable}
\pgfplotstabletypeset[
row predicate/.code={
\pgfplotstablegetelem{#1}{retrieved}\of{\rrranktable}
\ifnum\pgfplotsretval=1\relax
\else\ifnum\pgfplotsretval=5\relax
\else\ifnum\pgfplotsretval=10\relax
\else\ifnum\pgfplotsretval=20\relax
\else\pgfplotstableuserowfalse\fi\fi\fi\fi},
columns={retrieved, ocr-1, ocr-2,ocr-3,ocr-4, ocr-5, VLAD,rsvm,rr-5,rr-10,rr-15,rr-101,ocr-3-d, VLAD-d, rsvm-d, rr-15-d},
every head row/.style = {before row=\toprule, after row=\midrule},
every last row/.style={after row=\hline},
every row 0 column 10/.style={postproc cell content/.style={@cell content=\textbf{##1}}},
every row 4 column 10/.style={postproc cell content/.style={@cell content=\textbf{##1}}},
every row 9 column 10/.style={postproc cell content/.style={@cell content=\textbf{##1}}},
every row 19 column 10/.style={postproc cell content/.style={@cell content=\textbf{##1}}},
every row 0 column 15/.style={postproc cell content/.style={@cell content=\textbf{##1}}},
every row 4 column 15/.style={postproc cell content/.style={@cell content=\textbf{##1}}},
every row 9 column 14/.style={postproc cell content/.style={@cell content=\textbf{##1}}},
every row 19 column 14/.style={postproc cell content/.style={@cell content=\textbf{##1}}},
columns/rr-101/.style={ column type/.add={}{|}}
]{\rrranktable}  
\caption{Detailed retrieval rate values at specific points of curves as seen in other figures. ocr-$N$ represents the OCR-only curves of gram size $N$. VLAD represents VLAD-only curves. rsvm refers to OCR+VLAD combined results with learned weighting. rr-$K$ refers to the RANSAC-reranked curves based on the top $K$ on top of rsvm results. Columns suffixed with a '-d' refer to retrieval rates with all distractor data included. The best performing values are bolded in each row (based on un-rounded value). The distractor set results are bolded separately from the rest. }
\label{tbl:detailed_pnts}
\end{table*}

\subsection{Performance with Distractors}
\label{subsec:distractor_perf}
Finally, we evaluate our methods with the presence of over 100K distractors. We looked at how the additional distractor data affected individual feature performances, as well as with everything combined. 

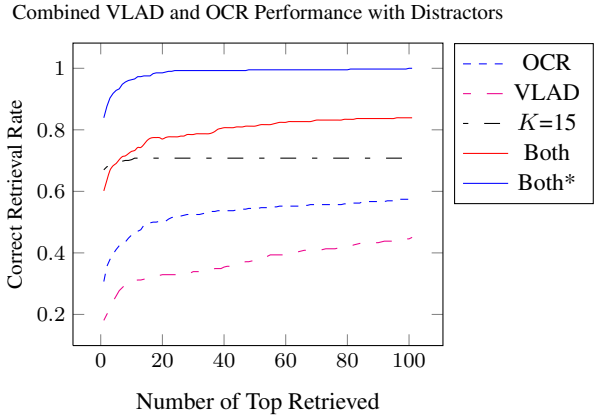
\begin{figure}
  \begin{center}
    \begin{tikzpicture}
      \pgfplotsset{small}
      \pgfplotstableread{data/ocr_ranks.dat}
      \ocrranktable
      \pgfplotstableread{data/vlad_ranks.dat}
      \vladranktable
      \pgfplotstableread{data/hybrid_ranks.dat}
      \hybridranktable
      \pgfplotstableread{data/rr_rank.dat}
      \rrranktable
      \footnotesize
      \begin{axis}[title=Combined VLAD and OCR Performance with Distractors,
        xlabel=Number of Top Retrieved,
        ylabel=Correct Retrieval Rate,
        ylabel near ticks,
        legend columns = 1,
        legend entries={OCR, VLAD, $K$=15, Both, Both*},
                legend style={font = \small},
        legend pos= outer north east]
        \addplot+[mark=none, dashed] table[x = retrieved,, y = ocr-3-d] from \ocrranktable;
        \addplot+[color=magenta, mark=none, dash pattern=on 3pt off 6pt on 6pt off 6pt] table[x = retrieved,, y = VLAD-d] from \vladranktable;
         \addplot+[color=black, mark=none, dash pattern=on 2pt off 6pt on 6pt off 6pt] table[x = retrieved,, y = rr-15-d] from \rrranktable;
        \addplot+[color=red, mark=none, smooth] table[x = retrieved,, y = rsvm-d] from \hybridranktable;
        \addplot+[color=blue, mark=none] table[x = retrieved,, y = rsvm] from \hybridranktable;
      \end{axis}
    \end{tikzpicture}
  \end{center}
\caption{Retrieval performance with distractors. Here, Both refers to OCR+VLAD, $K$=15 is OCR+VLAD+ransac reranking on top 15. We include the non-distracted curve (Both*) for ease of comparison. Results were plotted up to 101 retrievals (number of non-distractor images).}
\label{fig:OCR-VLAD-hybrid-dist}
\end{figure}

As with previous evalutations, we plot the retrieval rate against the size of the retrieval set in Fig.~\ref{fig:OCR-VLAD-hybrid-dist}. While all aspects took a significant hit in performance, the combined performance is now significantly better than that of VLAD or OCR alone as compared to the relatively tight gap between curves in Fig.~\ref{fig:OCR-VLAD-hybrid-rsvm}. Finally, we once again demonstrate that with an inexpensive ransac reranking of just the top 15 retrieved candidates still yields up to 7\% improvement for the accuracy of the top 1 result as seen in the last two columns of Table~\ref{tbl:detailed_pnts}.

Finally, we note that OCR was more robust to the effect of the added distractors than VLAD. At the top 1 retrieval, OCR dropped from 0.65 to 0.31, a difference of 0.34. VLAD on the other hand dropped from 0.68 to 0.18, a much greater drop of 0.5. This suggests that OCR can be more discriminative than VLAD overall on this specific data. Nevertheless, the fact that the combined performance is significantly better than either alone suggests that the types of mistakes they made were different enough such that combining was able to correct many of them.

\section{Conclusion}
We present a starting point for future research in large-scale book cover retrieval. As existing work in book cover retrieval is fairly limited and their datasets are lacking in realistically readily available author and title text annotatoions,  we  first augment an existing mobile-image based book cover dataset to make-up for these shortcomings. Building on the mobile book cover dataset provided by ~\cite{chandrasekhar2011stanford}, we first expand it with over 100K distractor cover images to emulate large scales, then include author and title information as a single string for each queryable cover (including the originally provided 101 covers).

Using our augmented dataset, we then demonstrated the general effectiveness of using text-based information in conjunction with other traditional BoW features such as VLAD. Because our query images are poorly conditioned and ill-suited for off-the-shelf OCR techniques, we demonstrate the use of character $N$-grams to robustly match against clean text annotations while using erroneous and noisy OCR output.

We recognize that many techniques we used can be easily improved upon and are far from comprehensive. However, our primary goal was to demonstrate the effectiveness of text-based information in realistic retrieval settings. Future work includes accelerating and improving the RANSAC/HOG re-ranking procedure, improving the robustness of our text-line extraction procedure, and trying new additional BoW features to further improve retrieval accuracy.

{\small
\bibliographystyle{ieee}
\bibliography{egbib}

\begin{thebibliography}{1}\itemsep=-1pt

\bibitem{chandrasekhar2011stanford}
V.~R. Chandrasekhar, D.~M. Chen, S.~S. Tsai, N.-M. Cheung, H.~Chen, G.~Takacs,
  Y.~Reznik, R.~Vedantham, R.~Grzeszczuk, J.~Bach, et~al.
\newblock The stanford mobile visual search data set.
\newblock In {\em Proceedings of the second annual ACM conference on Multimedia
  systems}, pages 117--122. ACM, 2011.

\bibitem{gomez2013multi}
L.~Gomez and D.~Karatzas.
\newblock Multi-script text extraction from natural scenes.
\newblock In {\em Document Analysis and Recognition (ICDAR), 2013 12th
  International Conference on}, pages 467--471. IEEE, 2013.

\bibitem{Joachims2006RSVM}
T.~Joachims.
\newblock Training linear svms in linear time.
\newblock In {\em Proceedings of the 12th ACM SIGKDD International Conference
  on Knowledge Discovery and Data Mining}, KDD '06, pages 217--226, New York,
  NY, USA, 2006. ACM.

\bibitem{Navarro00indexingmethods}
G.~Navarro, R.~Baeza-yates, E.~Sutinen, and J.~Tarhio.
\newblock Indexing methods for approximate string matching.
\newblock {\em IEEE Data Engineering Bulletin}, 24:2001, 2000.

\bibitem{shahab2011icdar}
A.~Shahab, F.~Shafait, and A.~Dengel.
\newblock Icdar 2011 robust reading competition challenge 2: Reading text in
  scene images.
\newblock In {\em Document Analysis and Recognition (ICDAR), 2011 International
  Conference on}, pages 1491--1496. IEEE, 2011.

\bibitem{smith2007overview}
R.~Smith.
\newblock An overview of the tesseract ocr engine.
\newblock In {\em ICDAR}, volume~7, pages 629--633, 2007.

\bibitem{Tsai2011Spine}
S.~S. Tsai, D.~Chen, H.~Chen, C.-H. Hsu, K.-H. Kim, J.~P. Singh, and B.~Girod.
\newblock Combining image and text features: A hybrid approach to mobile book
  spine recognition.
\newblock In {\em Proceedings of the 19th ACM International Conference on
  Multimedia}, MM '11, pages 1029--1032, New York, NY, USA, 2011. ACM.

\end{thebibliography}

\begin{thebibliography}{10}\itemsep=-1pt

\bibitem{bay2006Surf}
H.~Bay, T.~Tuytelaars, and L.~Van~Gool.
\newblock Surf: Speeded up robust features.
\newblock In {\em Computer Vision--ECCV 2006}, pages 404--417. Springer, 2006.

\bibitem{chandrasekhar2011stanford}
V.~R. Chandrasekhar, D.~M. Chen, S.~S. Tsai, N.-M. Cheung, H.~Chen, G.~Takacs,
  Y.~Reznik, R.~Vedantham, R.~Grzeszczuk, J.~Bach, et~al.
\newblock The stanford mobile visual search data set.
\newblock In {\em Proceedings of the second annual ACM conference on Multimedia
  systems}, pages 117--122. ACM, 2011.

\bibitem{chen2010building}
D.~M. Chen, S.~S. Tsai, B.~Girod, C.-H. Hsu, K.-H. Kim, and J.~P. Singh.
\newblock Building book inventories using smartphones.
\newblock In {\em Proceedings of the international conference on Multimedia},
  pages 651--654. ACM, 2010.

\bibitem{gomez2013multi}
L.~Gomez and D.~Karatzas.
\newblock Multi-script text extraction from natural scenes.
\newblock In {\em Document Analysis and Recognition (ICDAR), 2013 12th
  International Conference on}, pages 467--471. IEEE, 2013.

\bibitem{hariharan2012discriminative}
B.~Hariharan, J.~Malik, and D.~Ramanan.
\newblock Discriminative decorrelation for clustering and classification.
\newblock In {\em Computer Vision--ECCV 2012}, pages 459--472. Springer, 2012.

\bibitem{jegou2008hamming}
H.~Jegou, M.~Douze, and C.~Schmid.
\newblock Hamming embedding and weak geometric consistency for large scale
  image search.
\newblock In {\em Computer Vision--ECCV 2008}, pages 304--317. Springer, 2008.

\bibitem{jegou2010aggregating}
H.~J{\'e}gou, M.~Douze, C.~Schmid, and P.~P{\'e}rez.
\newblock Aggregating local descriptors into a compact image representation.
\newblock In {\em Computer Vision and Pattern Recognition (CVPR), 2010 IEEE
  Conference on}, pages 3304--3311. IEEE, 2010.

\bibitem{Joachims2006RSVM}
T.~Joachims.
\newblock Training linear svms in linear time.
\newblock In {\em Proceedings of the 12th ACM SIGKDD International Conference
  on Knowledge Discovery and Data Mining}, KDD '06, pages 217--226, New York,
  NY, USA, 2006. ACM.

\bibitem{malisiewicz2011ensemble}
T.~Malisiewicz, A.~Gupta, and A.~A. Efros.
\newblock Ensemble of exemplar-svms for object detection and beyond.
\newblock In {\em Computer Vision (ICCV), 2011 IEEE International Conference
  on}, pages 89--96. IEEE, 2011.

\bibitem{tfidf}
C.~D. Manning, P.~Raghavan, and H.~Sch\"{u}tze.
\newblock Introduction to information retrieval.
\newblock {\em Cambridge University Press 2008}, 2008.

\bibitem{MSER}
J.~Matas, O.~Chum, M.~Urban, and T.~Pajdla.
\newblock Robust wide baseline stereo from maximally stable extremal regions.
\newblock In {\em British Machine Vision Conference}, pages 384--393, 2002.

\bibitem{matsushita2011interactive}
K.~Matsushita, D.~Iwai, and K.~Sato.
\newblock Interactive bookshelf surface for in situ book searching and storing
  support.
\newblock In {\em Proceedings of the 2nd Augmented Human International
  Conference}, page~2. ACM, 2011.

\bibitem{Navarro00indexingmethods}
G.~Navarro, R.~Baeza-yates, E.~Sutinen, and J.~Tarhio.
\newblock Indexing methods for approximate string matching.
\newblock {\em IEEE Data Engineering Bulletin}, 24:2001, 2000.

\bibitem{philbin2007object}
J.~Philbin, O.~Chum, M.~Isard, J.~Sivic, and A.~Zisserman.
\newblock Object retrieval with large vocabularies and fast spatial matching.
\newblock In {\em Computer Vision and Pattern Recognition, 2007. CVPR'07. IEEE
  Conference on}, pages 1--8. IEEE, 2007.

\bibitem{shahab2011icdar}
A.~Shahab, F.~Shafait, and A.~Dengel.
\newblock Icdar 2011 robust reading competition challenge 2: Reading text in
  scene images.
\newblock In {\em Document Analysis and Recognition (ICDAR), 2011 International
  Conference on}, pages 1491--1496. IEEE, 2011.

\bibitem{shao2003zubud}
H.~Shao, T.~Svoboda, and L.~Van~Gool.
\newblock Zubud-zurich buildings database for image based recognition.
\newblock {\em Computer Vision Lab, Swiss Federal Institute of Technology,
  Switzerland, Tech. Rep}, 260, 2003.

\bibitem{smith2007overview}
R.~Smith.
\newblock An overview of the tesseract ocr engine.
\newblock In {\em ICDAR}, volume~7, pages 629--633, 2007.

\bibitem{Tsai2011Spine}
S.~S. Tsai, D.~Chen, H.~Chen, C.-H. Hsu, K.-H. Kim, J.~P. Singh, and B.~Girod.
\newblock Combining image and text features: A hybrid approach to mobile book
  spine recognition.
\newblock In {\em Proceedings of the 19th ACM International Conference on
  Multimedia}, MM '11, pages 1029--1032, New York, NY, USA, 2011. ACM.

\end{thebibliography}
}

\end{document}